# Experimental Visualization of Dispersion Characteristics of Backward Volume Spin Wave Modes

Sergey V. Gerus, Alexander Yu. Annenkov and Edwin H. Lock

(*Kotel'nikov Instutute of Radio Engineering and Electronics (Fryazino branch), Russian Academy of Sciences, Fryazino, Moscow region, Russia*)

Basing on the measurement of spatial spectra (spectra of wavenumbers), the dispersion characteristics of the first three modes of backward volume spin wave, propagating along the direction of a constant uniform magnetic field in a tangentially magnetized ferrite film, were visualized firstly. The study was carried out by microwave probing of spin waves with subsequent use of spatial Fourier analysis of the complex wave amplitude for a series of frequencies. It was found that every $m$-th mode of the backward volume spins wave can be split into $n$ satellite modes due to the existence of layers with similar magnetic parameters in ferrite film. It was found that satellites of the first mode of this wave are excited most effectively, while satellites of the third mode – least effectively, and the effectiveness of satellites excitation decreases as the number $n$ increases. It is found that the theoretical dispersion dependencies of the first three modes of the wave coincide well with the experimental dispersion dependencies of the satellite mode that are excited most effectively.

## 1. Introduction

It is well known that the only dipole spin wave (SW) that can propagate in a ferrite plate along the direction of a tangent homogeneous magnetic field is backward volume spin wave (BVSW), also called the backward volume magnetostatic wave since it was first described in magnetostatic approximation [1]. In the following years, the main characteristics of BVSW and various devices using this wave have been described in several monographs and review works (see [2-9] and references given in these works), as well as in the number of recent papers [10-17], in which new results concerning BVSW were obtained.

From the publication of work [1] to the twenty-first century, researchers were mainly interested by the BVSW dispersion dependence, the amplitude-frequency characteristic of BVSW transfer coefficient and the delay time of BVSW between two transducers perpendicular to the external magnetic field $\mathbf{H}_0$ [1][1]. That is, the characteristics of the BVSWs with collinear orientation of wave vector $\mathbf{k}$ and group velocity vector $\mathbf{V}$ were investigated to construct various filters, delay lines and other devices for analogue signal processing at microwave frequencies.

---

[1] Including cases when transducers of various shapes were used or various inhomogeneities were placed between them – a lattice of conductive strips, etched grooves, etc.

Theoretical and experimental studies of BVSW characteristics with noncollinear orientation of the vectors **k** and **V** really started with the works [10, 11]. In particular, the microwave field distribution of the BVSW propagating at an arbitrary angle to vector $\mathbf{H}_0$ has been investigated theoretically and the instantaneous vector lines patterns of magnetic field for the first mode of BVSW have been calculated [10]. In addition, it has been shown theoretically and experimentally that excitation of BVSW by a linear transducer which is not perpendicular to the $\mathbf{H}_0$ vector gives rise to appearance of two waves characterized by oppositely directed wave vectors and different magnetic potential distributions[2] across the thickness of ferrite plate, that course to the significant difference between amplitudes of these two excited waves[3] [11]. The condition under which two reflected beams appear and the condition under which a negative reflection[4] occurs have also been described for the geometry where BVSW incidents on the straight edge of ferrite film [11, 12]. As a result of further theoretical studies of the properties of BVSW, it was proved that propagation of this wave, like the propagation of the surface SW, is characterized by the presence of wave vector cut-off angles [15, 17]. In addition, it was found that an extremum point[5] appears on the distribution of magnetic potential amplitude for the first BVSW mode at a certain orientation of its wave vector, and the coordinate of this point corresponds to the one of ferrite plate surfaces [17]. It has also been found that the degree of wave non-reciprocity, defined as the ratio of normalized amplitudes of the magnetic potential on a ferrite plate surface for two waves with oppositely directed wave vectors, depends significantly on the orientation of these vectors with respect to the vector $\mathbf{H_0}$ [17].

Rather recently, the diffraction properties of the first mode of the BVSW have been investigated theoretically and experimentally [14, 16]. In particular, based on theoretical results obtained in [18], for the first BVSW mode it was calculated the wave vector orientations [14], at which the angular width of the wave beam is zero and the wave is characterized by super-directional propagation which was then discovered experimentally in [16].

The description of BVSW without magnetostatic approximation has shown that the electromagnetic wave, named in [1] as BVSW, has all six microwave electromagnetic field components (three magnetic and three electric) both in the ferrite layer and in the adjacent half-spaces: that is, both E-wave and H-wave, which are coupled to each other due to the presence of ferrite plate, arise in adjacent half-spaces as a result of the satisfaction of electrodynamic boundary conditions [13]. As a result of these studies the exact dispersion equation for this wave was obtained (for the case of wave propagation along the vector $\mathbf{H_0}$) and it was shown that the distribution of all its microwave components across the thickness of the ferrite plate is not purely trigonometric (which follows from [1]) but is a sum of exponential and trigonometric functions, and the cross components of the wave number included in trigonometric and exponential functions have different values [13]. Obviously, this fact means that BVSW is not a purely volume wave and gives the reason to rename this wave. However, over the past time, the term BVSW has become widespread and we think that, to avoid confusion of terms, it would not be appropriate to use another term with respect to this wave just because new its properties have been discovered or other equations have been used for its description: nevertheless, this wave is an objective reality and its name should not depend on the way of its theoretical description.

---

[2] This means that BVSW is characterized by non-reciprocal properties which were previously described only for the surface SW [1].

[3] An exception is the case where both waves propagate parallel to the external magnetic field vector in opposite directions. Only in this case both waves have the same magnetic potential distribution across the thickness of ferrite plate and, therefore, are excited with the same amplitudes.

[4] That is, when the incident and reflected beams are on the same side of the boundary normal.

[5] This point is the only extremum point on the distribution of magnetic potential amplitude for the first mode of BVSW. It should also be noted that the coordinate of this point gradually shifts towards the middle of ferrite plate as the angle between the BVSW wave vector and vector $\mathbf{H_0}$ increases.

It should be noted that all the studies listed above are mainly devoted to the first mode of BVSW. The characteristics and properties of the higher modes of BVSW are still almost not investigated: only the dispersion dependences and magnetic potential distribution of higher modes of BVSW have been calculated for waves propagating along vector $\mathbf{H_0}$ [1] and for waves propagating in arbitrary directions [10, 17], and it was also shown that all modes of BVSW are characterized by the same cut-off angles of the wave vector [15]. However, the properties and characteristics of the higher modes of BVSW have not yet been studied experimentally.

Filling this gap, we present below an experimental study of the spatial spectrum of BVSW modes which are excited in a tangentially magnetized ferrite plate by a linear transducer perpendicular to the vector $\mathbf{H_0}$.

## 2. Experimental setup and method for the probing of spin waves

At present, measurement of SW characteristics is generally carried out using the method of Brillouin light scattering on SW (see, for example, [19, 20]) or the method of SW probing [16, 21, 22], which allow to obtain visualized patterns describing the amplitude and the phase distribution of SW in the plane of ferrite film or ferrite structure. To carry out the planned experiments, there was used SW probing method, which, as will be shown below, allowed to decide all experimental problems.

The method of SW probing emerged from the step-by-step development of the moving antenna method[6] in which a receiver transducer was moved over a ferrite

---

[6] It is the first method used to measure the dispersion characteristics of SWs in the 1970s - 1990s

structure surface along the SW path to measure the phase shift of SW at certain distance and then to find the wave number at fixed frequency (see, for example, [23, 24]). Subsequently, both the measurement method and the experimental setup were improved significantly (see Fig. 1).

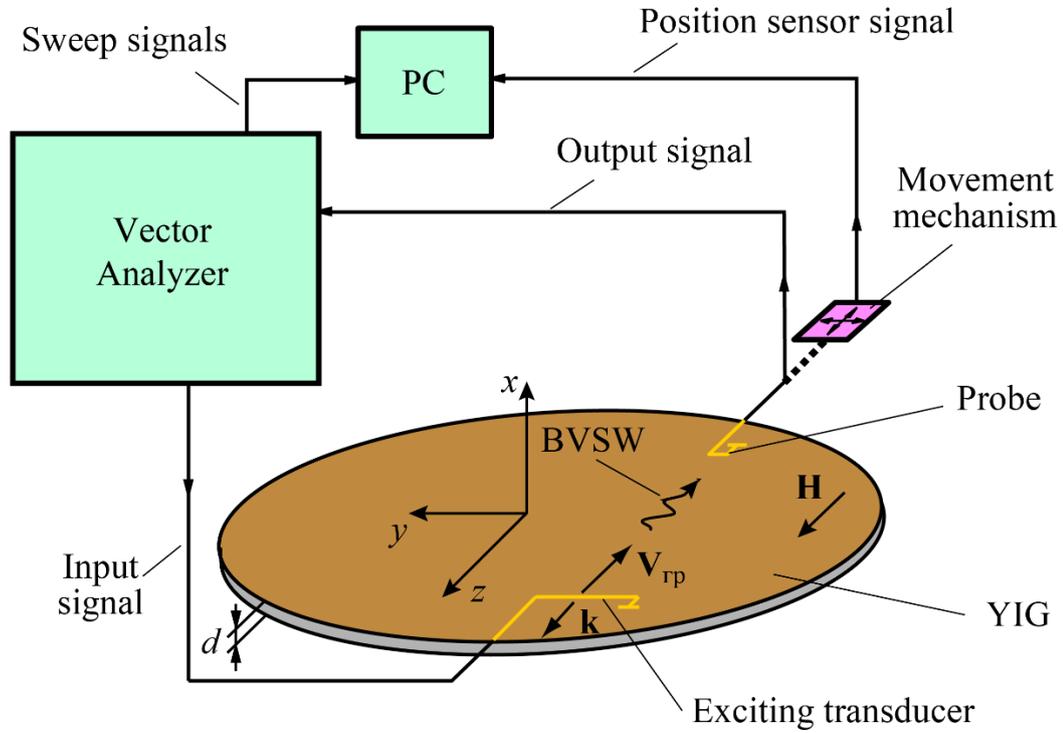

Fig. 1. Simplified scheme of experimental setup.

The microwave probe, having form of a loop of thin gold-plated tungsten wire with an aperture of ~0.5 mm, began to be used in place of the receiving transducer (which was previously identical to the exciting transducer) in the experimental setup. Both the probe and the exciting transducer were equipped with position sensors, could rotate around the axis normal to the ferrite structure plane, and could move freely over the structure surface along $y$ and $z$ axes by means of movement mechanism. SW characteristics were measured in the following way: for a number of fixed coordinates $y$, a continuous probe movement over the structure surface along $z$ axis was carried out with simultaneous digitization of both the instantaneous $z$ coordinate values and the complex amplitude of the microwave transfer coefficient between transducers [21]. Since it is SW that carries the microwave signal in ferrite structure, then the probe, in fact, measures the complex amplitude of SW during its movement. As a result of processing the experimental data coming from the vector analyzer into the computer, it was possible to obtain visualized patterns of the amplitude and phase distribution for the studied SW in its propagation area at a fixed frequency $f = f_{\text{const}}$ (see, for example, the patterns presented in [16, 22]).

Then this method was improved again to obtain visualized wave distribution patterns over a wide frequency range during a single probe pass along the surface of the structure. For this purpose, during slow movement of the probe over the surface of the structure along $z$-axis a fast sawtooth change of the vector analyzer frequency was carried out simultaneously in a certain frequency range $f_{start} < f < f_{end}$, and the probe shift during the period of change of sawtooth voltage was very small.

As a result of this improvement, it was possible to measure the distribution of the complex transfer coefficient $K(y, z, f)$ as a function of the frequency $f$ and the $y$, $z$ coordinates along the structure surface.

### 3. Method for determining the spatial spectrum of spin waves

By processing the $K(y, z, f)$ distribution in different ways, a number of SW characteristics can be obtained. For example, using a discrete Fourier transform of the $K(y, z, f)$ distribution along the wave vector direction normal to the phase fronts[7], one can obtain the dependence of the Fourier amplitude $A$ on the wave number $k$ for each frequency $f$ from the frequency range $f_{start} < f < f_{end}$. Obviously, the positions of the maxima on the $A(k)$ dependence correspond to the spin wave numbers exciting at a certain frequency $f$ in experiment. Thus, in general, the dependence $A(k)$ is a spatial spectrum or wave numbers spectrum, the maxima of which correspond to the excited wave numbers for SW, or various SW modes, or any other waves that can excite and receive the used transducer and probe. It is clear that the sum of the maxima of all spatial spectra at the studied frequencies $A(k, f)$ represent the dispersion relations of observed waves. It should be noted, however, that in order to successfully use the described method to find the spatial waves spectrum $A(k, f)$ in ferrite structures, the experiments must be carried out with taking into account the well-known feature of Fourier analysis. In particular, it is known from the theory of spectral analysis of signals based on the Fourier transform that the longer the duration of a pulse, the narrower its spectrum (see, for example, §2.2. in [25]).

Since we use spatial Fourier analysis of the $K(y, z, f)$ distribution, the following statement is valid: the greater is the distance at which the $K(y, z, f)$ distribution is measured, the narrower are the maxima of the investigated spatial wave spectrum $A(k, f)$ and, therefore the greater are the accuracy and resolution one can obtain in the experiment to determine the spatial wave spectrum. This statement has been recently confirmed in experimental studies of SW characteristics [26].

---

[7] It is assumed that in experiment one studies a sufficiently wide wave beam, in the middle of which the phase fronts are straight lines, i.e., the wave is practically like a homogeneous plane wave. It is clear that in the experiment, more complex distributions of $K(y, z, f)$ may occur, when the phase fronts are not straight lines(such fronts often appear in diffraction patterns). However, a discussion of such complex distributions is beyond the framework of this paper.

#### 4. Spatial spectrum of spin waves propagating along the vector $\mathbf{H}_0$

The spatial spectrum of the spin wave was studied on a YIG film grown by liquid phase epitaxy on a 0.5 mm thick gallium gadolinium garnet (GGG) substrate. The YIG film, having a diameter of 76 mm, a thickness of $d = 39$ μm and a saturation magnetization of $4\pi M_0 = 1750$ G, was magnetized to saturation by a tangential homogeneous magnetic field of $H_0 = 483 \pm 5$ Oe. A 10 mm long linear transducer was used to excite SW. The transducer was oriented perpendicular to vector $\mathbf{H}_0$ and positioned near the edge of the film by such means that the normal passing through the center of the transducer did not pass through the center of the film circle (Fig. 1). This orientation of the transducer provided excitation of any types of SW (including BVSW modes) with wave vectors directed along $z$-axis, and asymmetric location of transducer respect to the film center prevented reflection of the waves from the near film edge back towards the transducer and interference of these waves with the waves running towards the far film edge, where the probing area was located.

To describe the obtained results, a Cartesian coordinate system is used in which the $x$-axis is perpendicular to the film plane and the $z$-axis is directed along the vector $\mathbf{H}_0$, with coordinate $z = 0$ corresponding to the projection of exciting transducer middle onto the film plane.

The spatial distribution of the complex BVSW amplitude measured during probe movement parallel to the magnetic field $\mathbf{H}_0$ is shown in Fig. 2 for arbitrarily frequency value. As can be seen from Fig. 2, the oscillation amplitude decreases quite rapidly with increasing distance between the probe and the excitation transducer, that is related mainly with BVSW beam diffraction divergence (rather than losses), which is quite large for this geometry of wave excitation [17, 18]. The subtle distortions of the oscillation curves show that besides the most intensely excited first mode BVSW, there are modes with higher values of the wave number. This is also confirmed by Fig. 3, which shows the wave vectors spectrum $A(k)$ obtained from a Fourier analysis of the measured complex amplitude of BVSW for an arbitrary fixed frequency. The relative magnitude of the maxima on the dependence $A(k)$ differs greatly and characterizes the intensity of BVSW modes excitation: the greatest amplitude has a maximum located near the theoretical value of the wave number $k_1^t(f_1)$ for the first BVSW mode (Fig. 3a), and the smallest – a maximum located near the theoretical value of the wave number $k_3^t(f_1)$ for the third BVSW mode (Fig. 3c).

Note that the complex amplitude of BVSW was measured when the probe was moving in the negative $z$-axis direction, i.e., in the wave energy transfer direction described by the group velocity vector. However, from the Fourier analysis we obtained positive values of wave numbers $k_1$, $k_2$ and $k_3$ for BVSW modes, that corresponds to the positive orientation of wave vectors **k₁**, **k₂** and **k₃** with respect to the $z$-axis. Thus, opposite orientations of wave vectors and corresponding group velocity vectors confirm that we are dealing with the backward waves.

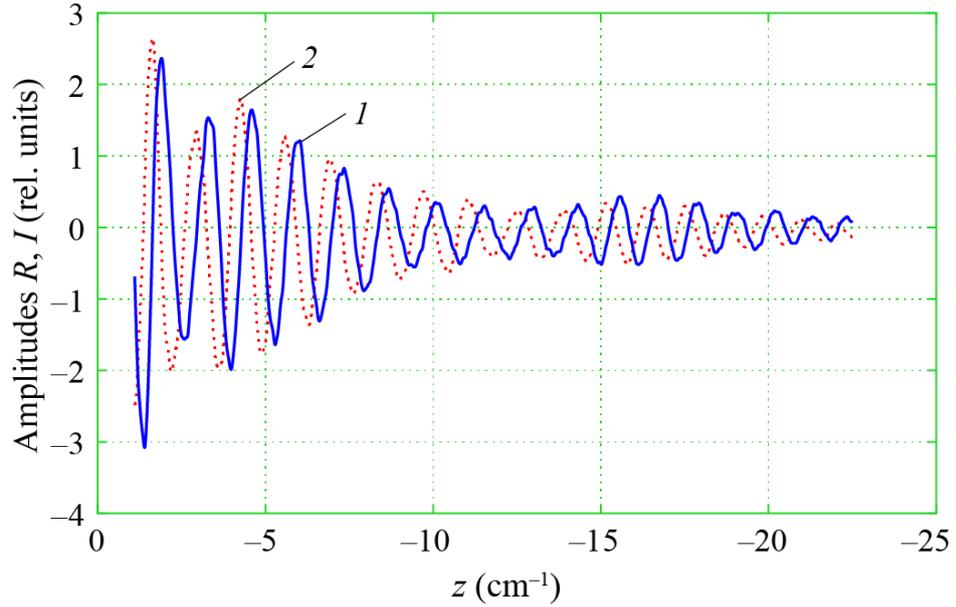

Fig. 2. Real $R$ (*1*) and imaginary $I$ (*2*) parts of the complex amplitude as a function of the $z$-coordinate when microwave energy is transferred between the excitation transducer and the probe by means of BVSW at the frequency $f_1 = 2808$ MHz.

Since the maxima of each spectrum $A(k)$ correspond to the wave numbers of the BVSW modes at a fixed frequency $f$ (e.g., in Fig. 3, $f = f_1$), so obviously, by uniting all these spectra for different frequencies, we obtain the surface $A(k, f)$, on which the maxima of the spectra correspond to the dispersion dependences $f(k)$ for BVSW modes excited in the experiment. Such a surface is shown in Fig. 4.

The darkest regions of the spectrum $A(k, f)$ correspond to amplitudes $A$ that are close to zero, while the lightest regions correspond to the wave numbers maxima obtained by Fourier analysis. The set of these maxima in Fig. 4 represents the visualized dispersion dependencies $f(k)$ of all waves observed in this experiment. In addition, the white points in Fig. 4 correspond to the several frequency and wave number values describing theoretical dispersion dependences of the first, second, and third modes of BVSW in magnetostatic approximation for the experimental parameters. As can be seen from Fig. 4, the theoretical points generally agree well with the experimental curves.

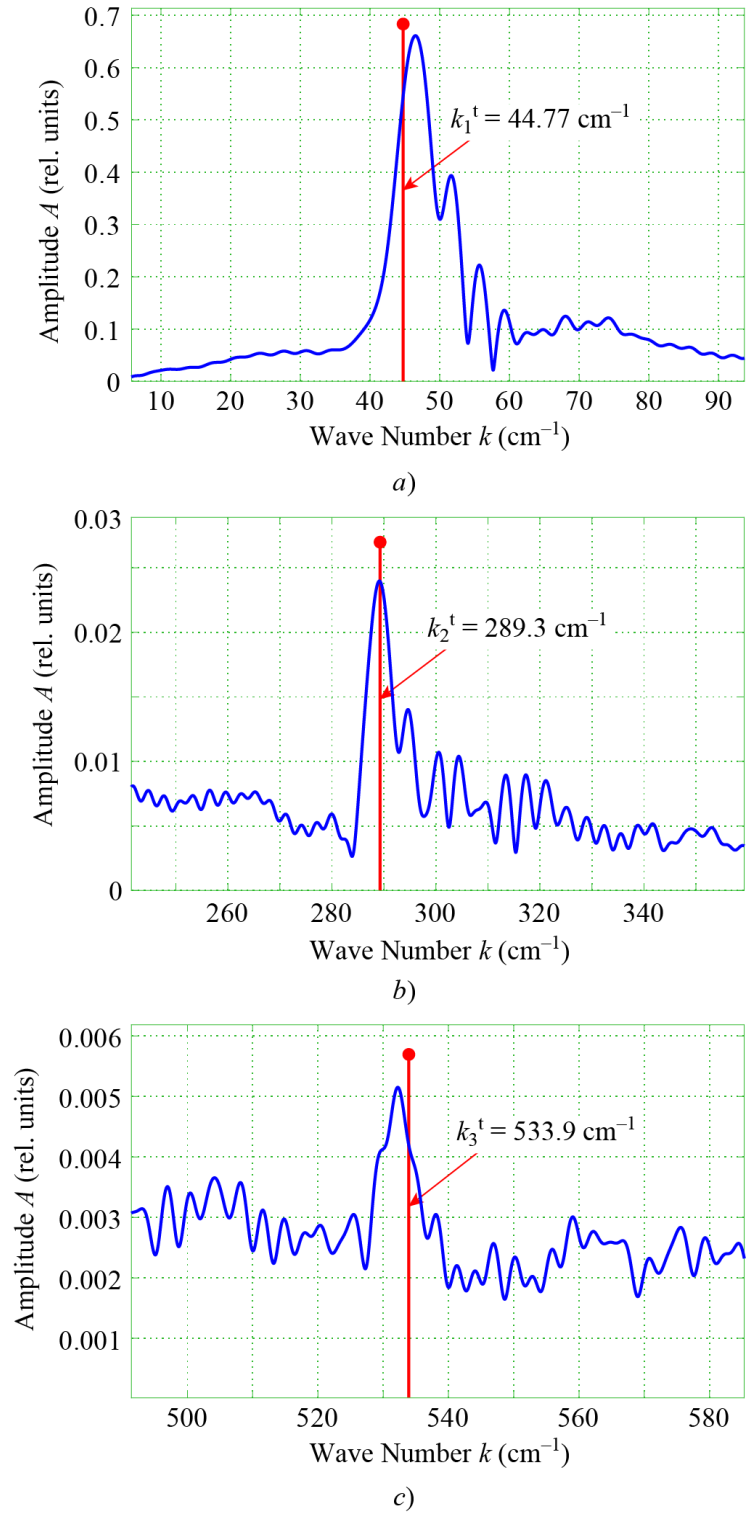

Fig. 3. Fourier amplitude versus wave number $A(k)$ at the frequency $f_1 = 2808$ MHz near the theoretically calculated wavenumber values $k_1^t(f_1)$ $k_2^t(f_1)$ corresponding to the vertical lines for the first, second and third BVSW modes (a, b and c, respectively).

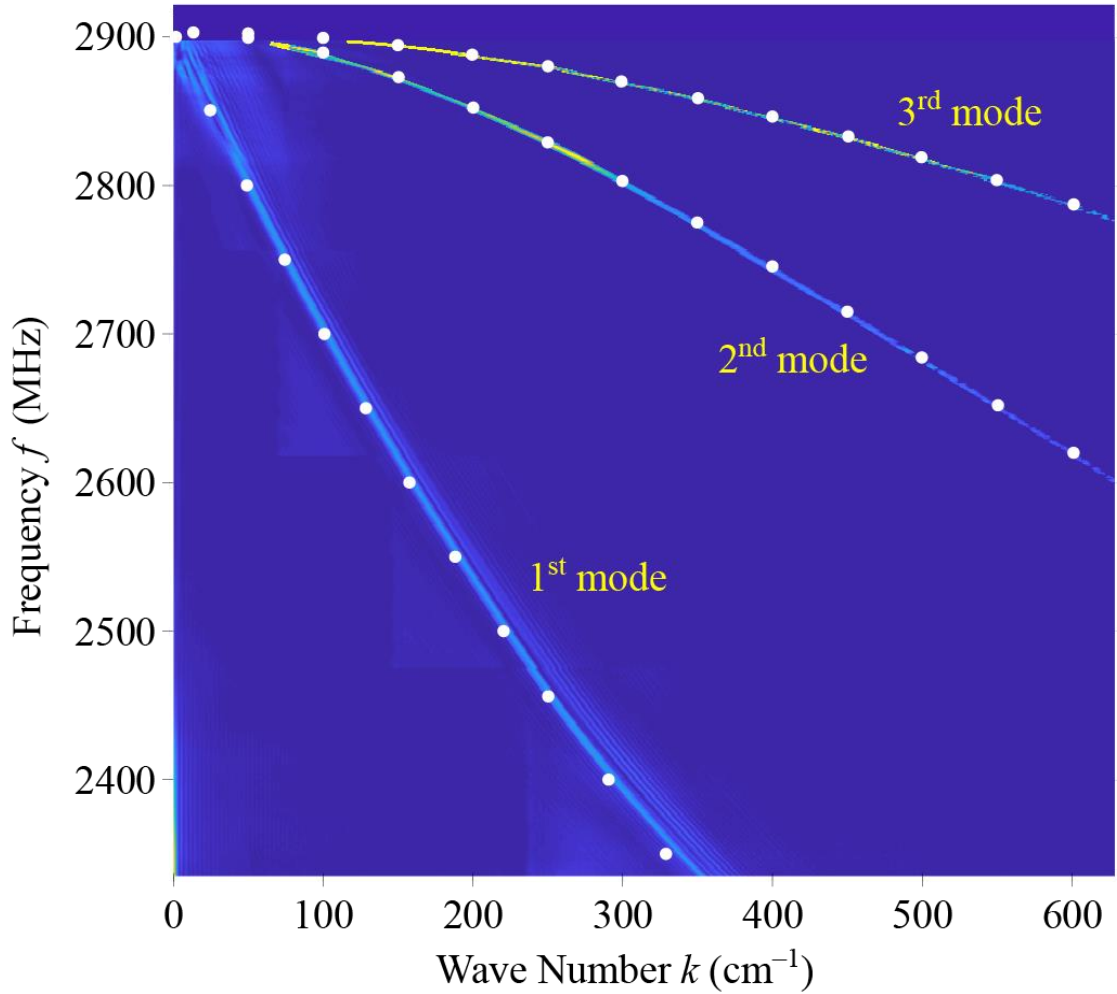

Fig. 4. Graphical illustration of the experimental surface $A(k, f)$, visualizing the dispersion dependences $f(k)$ for the first three BVSW modes. The white circles correspond to the theoretical dispersion dependences for these three BVSW modes (mode numbers are shown near the curves).

However, it should be noted that the maxima of the experimental spectrum $A(k, f)$ visualize in Fig. 4 much more dispersion dependences then follows from the theory [1]. These additional dispersion dependences are especially well seen near the theoretical dispersion dependence for the first mode of BVSW (Fig. 4). This fact is also clearly seen in Fig. 3a: near the theoretical value of the wave number $k_1^t(f_1)$ for the first BVSW mode, there is not one experimental maximum, but $n$ closely located maxima at $k_{1n}^{\exp}(f_1)$ values.

It may be assumed that series of closely located dispersion dependences observed experimentally are explained by splitting of the first BVSW mode into satellite modes[8], the excitation of which is caused by the stratification of YIG film in the process of its growth with formation of several layers having similar parameters of saturation magnetization or growth anisotropy.

---

[8] Apparently, the distribution of the magnetic potential of all modes-satellites over the thickness of the ferrite film is like a similar distribution for the first mode of the BVSW according to the theory [1]: that is, the distribution of modes-satellites is described by an odd, sinusoidal, centrally symmetric relative to the middle of the film wave function (see, for example, [17, Figure 2a, curve 1]).

Fragments of the dispersion dependences in which satellite modes of the first BVSW mode are clearly distinguishable are shown in Fig. 5 in a more detailed scale. These dispersion dependences correspond to the maxima of the $A(k)$ spectra (such as are seen in Figure 3a) for the certain frequency interval. Bold lines in Fig. 5 identify parts of the dispersion curves with maxima $A(k)$ having highest values. For example, in Fig. 3a this is the left maximum, which is significantly larger than the others. In Fig. 4 such curves, to which the highest maxima of $A(k)$ correspond, are shown by the lightest lines.

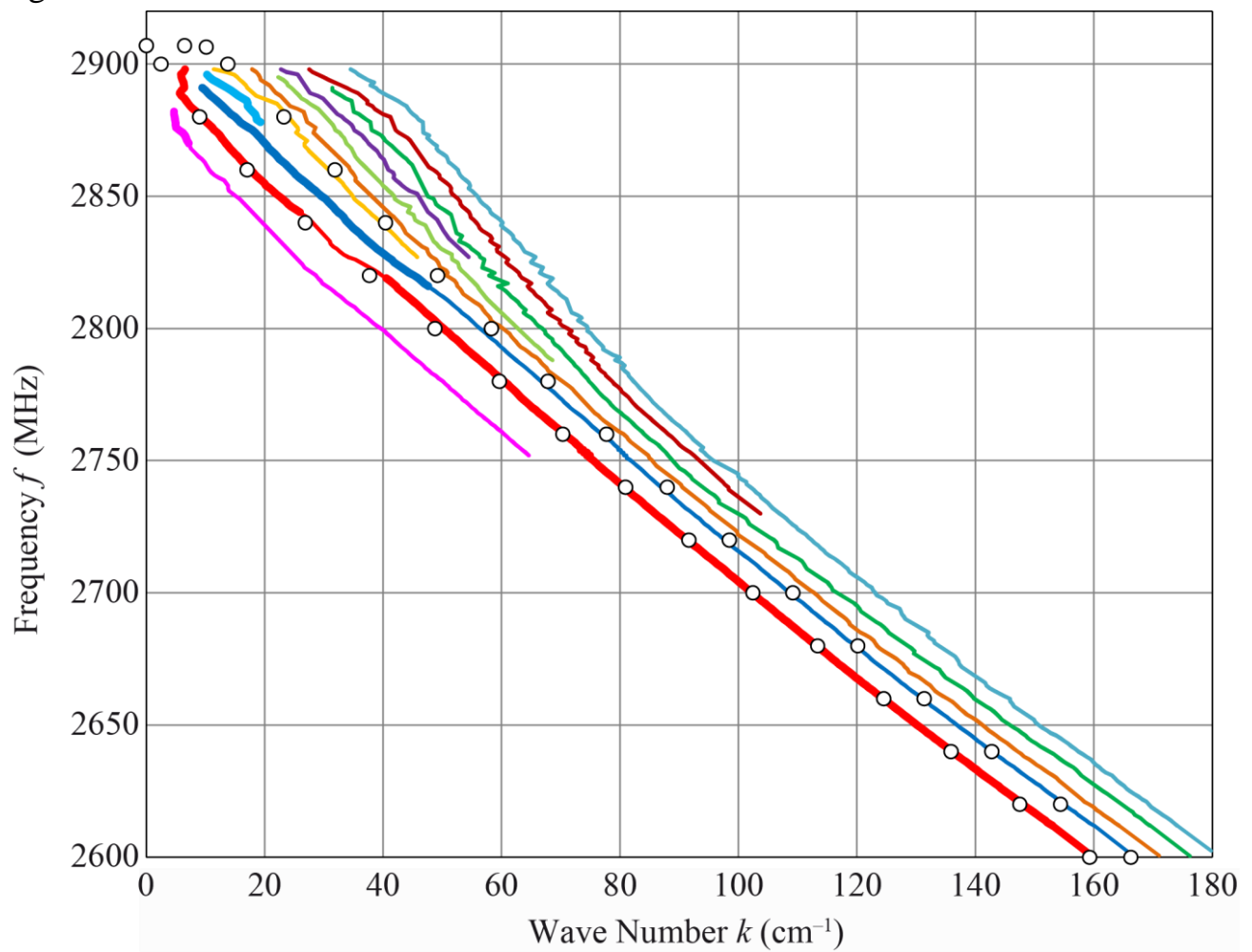

Fig. 5. Experimental dispersion dependences $f(k)$ for the first BVSW mode in a more detailed scale (dependences $f(k)$ correspond to the dependences $A(k, f)$, in Fig. 4). The circles correspond to the theoretical dispersion dependences of this mode in the YIG1–GGG–YIG2 structure.

The presence of layers in garnet ferrite films was observed earlier in the study of films with cylindrical domains [27]: when the layers were chemically etched off successively, the domain structure was changed by jumps. However, there is still no complete understanding of the cause of the formation of such layers. There are several hypotheses. It is supposed that during epitaxial growth of the ferrite film on a non-magnetic gadolinium-gallium substrate elastic stresses arise and build up due to some lattice mismatch, and these stresses at some moment decrease by jump due to the restructuring of the ferrite film lattice. In addition, during the growth process a subtle, but significant for the growth, change in the melt composition may occur. A combination of these mechanisms is possible too.

In addition, by analyzing Fig. 5, one can notice that in the range of small wave numbers $k < \sim 70$ cm$^{-1}$ (Fig. 5a) there are observed dispersion curves, which are absent in the region of higher values of $k$. The appearance of these curves is evidently related to the interaction between BVSW modes, which propagate in different YIG films grown on opposite substrate surfaces of $h = 0.5$ mm thickness. Indeed, since along the $x$-axis, normal to YIG plane, the microwave fields decrease according to the exp(-$kx$) law, so at $k = 20$ cm$^{-1}$ and $x = h$ the exponent index becomes equal to unity, which shows the possibility of interaction between these BVSW modes and the possibility of excitation of both modes by the transducer located on one of the film surfaces (a similar effect occurring for surface SW was described earlier in [21]).

To verify the mentioned interaction the dispersion dependences of the first BVSW modes, which propagate in two ferrite layers separated by substrate of thickness $h$, were calculated. In calculations it was assumed that the layers had the same thickness of 39 μm and saturation magnetizations of 1852.1 and 1874.3 G. The calculated dispersion dependences are plotted by dots in Fig. 5, which shows that the curves corresponding to the two almost identical ferrite layers differ most at $f > \sim 2750$ MHz and $k < \sim 70$ cm$^{-1}$, while at $f < \sim 2750$ MHz and $k > \sim 70$ cm$^{-1}$ these curves gradually approach each other by a distance of $\sim 7$ cm$^{-1}$.

Since it was assumed that several ferrite layers are formed on each side of the substrate, so a series of dispersion dependences for satellite modes of the first BVSW mode in the form of a stretched loop is observed in the frequency range $\sim 2750$ MHz $< f < \sim 2900$ MHz. With increasing wave number at $k > \sim 70$ cm$^{-1}$ the interaction between BVSW modes localized in different layers decreases, which leads to weakening of the second and subsequent experimental maxima observed in Fig. 3.

Calculations have shown that closely located dispersion dependences of BVSW modes do not arise for the geometry of contacting or closely located ferrite layers: a certain gap (or non-magnetic substrate) between layers with a thickness considerably less than the gap thickness is necessary for such dependencies to occur. However, a detailed study of this phenomenon is beyond the framework of this paper.

# 5. Summary

The spatial spectrum of waves propagating in a tangentially magnetized ferrite film along the direction of a constant homogeneous magnetic field is investigated experimentally. The study was carried out by means of microwave probing with the following use of Fourier analysis to visualize the spatial spectrum of the waves. As a result, it was found that the experimental spatial wave spectrum for this excitation geometry is significantly different from the theoretical BVSW spectrum calculated according to the theory [1]. In particular, it was found that each $m$-th mode of the BVSW predicted by the theory [1] can additionally split into $n$ satellite modes. It was found that the satellites of the first BVSW mode are excited most efficiently, while the satellites of the third BVSW mode are excited least efficiently, and the excitation efficiency of satellites decreases with the increasing of number n. The greatest number of mode satellites, about seven, was observed for the first BVSW mode, whose theoretical dispersion dependence matched well with the experimental dependence of its first satellite, which was most efficiently excited. The splitting of the second BVSW mode into satellite modes was inefficient, and the splitting of the third BVSW mode was practically not observed due to inefficient excitation of these modes, so the theoretical dispersion dependences of the second and third BVSW modes coincided well with the experimental dependences, which as a rule corresponded to a single maximum obtained by Fourier analysis (one may suppose that only one satellite mode was effectively excited for both the second and the third BVSW modes).

Also, the interaction between BVSW modes, which propagate in YIG films grown on opposite substrate surfaces, was observed experimentally for the wave numbers $k < \sim 70\ \mathrm{cm}^{-1}$. It was found that this interaction leads to a notable distortion of dispersion dependences for satellite modes of the first BVSW mode.

It can be assumed that BVSW modes split into satellite modes due to stratification of the ferrite film into several layers with the same magnetic parameters during epitaxial growth. The observed effect of splitting of the first BVSW mode into satellites can be used in practice to study the parameters of grown ferrite films.

## Funding

# References

1. Damon R. W. and Eshbach J. R., J. Phys. Chem. Solids**, 19** (1961) 308.

2. Pizzarello F. A., Collins J. H., Coerver L. E., J. Appl. Phys., **41** (1970) 1016.

3. Adam. J. D., Daniel M. R., IEEE Trans. on Magnetics, **MAG-17** (1981) 2951.

4. Castera J.-P., J. Appl. Phys., **55** (1984), 2506.

5. Danilov V. V., Zavislyak I. V., and Balinskii M. G., Spin Wave Electrodynamics (Libid', Kiev) 1991).

6. Vashkovskii A. V., Stalmakhov V. S., Sharaevskii Y. P., Magnetostatic Waves in Microwave Electronics (Saratov Gos. Univ., in Russian) 1993.

7. Gurevich A. G. and Melkov G. A., Magnetization Oscillations and Waves (CRC, Boca Raton, Fl.) 1996).

8. Stancil D.D. and Prabhakar A., Spin Waves: Theory and Applications (Springer Science + Business Media, New York) 2009.

9. Demokritov S.O. and Slavin A.N. (Editors), Magnonics: from Fundamentals to Applications vol. 125 (Springer-Verlag, Berlin) 2013.

10. Annenkov A. Yu. and Gerus S. V., Zh. Tekh. Fiz. **69** (1999) 82.

11. Vashkovsky A. V. and Lock E. H., Phys.-Usp. **49** (2006) 389.

12. Lock. E.H., Physics-Uspekhi**, 51** (2008) 375.

13. Vashkovskii A. V. and Lokk E. G., J. Commun. Technol. Electron. **57** (2012) 490.

14. Lokk E. G., J. Commun. Technol. Electron. **60** (2015) 97.

15. Lokk E. G., J. Commun. Technol. Electron. **63** (2018) 915.

16. Annenkov A.Yu., Gerus S.V., Lock E.H. EPJ Web of Conf. **185** (2018) 02006.

17. Lokk E. G., J. Commun. Technol. Electron. **65** (2020) 265.

18. Lock E. H., Phys.-Usp., **55** (2012) 1239.

19. Kruglyak V. V., Demokritov S. O., Grundler D., J. Phys. D. **43** (2010) 264001.

20. Sadovnikov A.V. Odintsov S. A., Beginin E. N. et al., Phys. Rev. B. **96** (2017) 144428.

21. Annenkov A.Yu., Gerus S.V., J. Commun. Technol. Electron. **57** (2012) 519.

22. Annenkov A.Yu., Gerus S.V., Lock E.H., EPL (Euro Phys. Lett.) **123** (2018) 44003.

23. Zubkov V. I., Lokk E. G., Shcheglov V. I., Sov. J. Commun. Technol. Electron. **34** (1989) 1381.

24. Zubkov V. I., Lokk E. G., Nam B. P. et al., Tech. Phys**, 34** (1989) 115.

25. Baskakov S.I., Radio circuits and signals (Moscow: Higher School) 1983.

26. Gerus S.V., Lock E.H., Annenkov A.Yu., J. Commun. Technol. Electron. **66** (2021) 1378.

27. Avaeva I.G., Kravchenko V.B., Lisovsky F.V. et al., Microelectronics **7** (1978) 444.